\documentclass[showpacs,reprint]{revtex4-1}
\usepackage{amsfonts}
\usepackage{amsmath}
\usepackage{amssymb}
\usepackage{graphicx}

\providecommand{\U}[1]{\protect\rule{.1in}{.1in}}
\providecommand{\U}[1]{\protect\rule{.1in}{.1in}}
\providecommand{\U}[1]{\protect\rule{.1in}{.1in}}

\ifx\pdfoutput\relax\let\pdfoutput=\undefined\fi
\newcount\msipdfoutput
\ifx\pdfoutput\undefined\else
\ifcase\pdfoutput\else
\ifx\paperwidth\undefined\else
\ifdim\paperheight=0pt\relax\else\pdfpageheight\paperheight\fi
\ifdim\paperwidth=0pt\relax\else\pdfpagewidth\paperwidth\fi
\fi\fi\fi
\ifx\pdfoutput\relax\let\pdfoutput=\undefined\fi
\newcount\msipdfoutput
\ifx\pdfoutput\undefined\else
\ifcase\pdfoutput\else
\ifx\paperwidth\undefined\else
\ifdim\paperheight=0pt\relax\else\pdfpageheight\paperheight\fi
\ifdim\paperwidth=0pt\relax\else\pdfpagewidth\paperwidth\fi
\fi\fi\fi

\begin{document}

\title{Crucial role of decoherence for electronic transport in molecular
wires: Polyaniline as a case study.}
\author{Carlos J. Cattena,$^{1}$ Ra\'{u}l A. Bustos-Mar\'{u}n,$^{1,2}$ and
Horacio M. Pastawski$^{1}$}

\affiliation{$^{1}$Instituto de F\'{\i}sica Enrique Gaviola (CONICET-U.N.C.) and Facultad
de Matem\'{a}tica Astronom\'{\i}a y F\'{\i}sica, Universidad Nacional de
C\'{o}rdoba, Ciudad Universitaria, C\'{o}rdoba, 5000, Argentina} 
\affiliation{$^{2}$Facultad de Ciencias Qu\'{\i}micas, Universidad Nacional de C\'{o}rdoba,
Ciudad Universitaria, C\'{o}rdoba, 5000, Argentina.}


\begin{abstract}
In this work we attempt to elucidate the nature of conductivity in polymers
by taking the acid-base doped polyaniline (PAni) polymer. We evaluate the
PAni conductance by using realistic \textit{ab initio} parameters and
including decoherent processes within the minimal parametrization model of
D'Amato-Pastawski. In contrast to general wisdom, which associates the
conducting state with coherent propagation in a periodic polaronic lattice,
we show that decoherence can account for high conductance in the strongly
disordered bipolaronic lattice. Hence, according to our results, there is no
need of considering a mix model of \textquotedblleft
conducting\textquotedblright\ polaronic lattice islands separated by
\textquotedblleft insulating\textquotedblright\ bipolaronic lattice strands
as is usually assumed for PAni. We find that without dephasing events, even
very short strands of bipolaronic lattices are not able to sustain
electronic transport. We also include a discussion of specific mechanisms
that should be involved in decoherence rates of PAni and relate them with
Marcus-Hush theory of electron transfer.
\end{abstract}

\maketitle

\section{Introduction}

In the late 70's, Alan MacDiarmid, Alan Heeger and Hideki Shirakawa led the
investigations which put conducting polymers at the center stage by
unraveling the transition from insulator to metal upon doping of
polyacetilene \cite{Heeger-Nobel}. The following decades, these materials
encountered numerous technological applications \cite{Applications,Conwell}.
The novelty of polyacetilene's physical properties, e.g. transport through
solitonic excitations \cite{SuSchriefferHeeger}, made it the most
intensively studied conductive polymer. However, the interest then shifted
to polyanilines \cite{electroactive-synt-met} and related compounds because
they are inexpensive, stable and easy to made.

\begin{figure}[tbp]
\begin{center}
\includegraphics[width=3.0in]{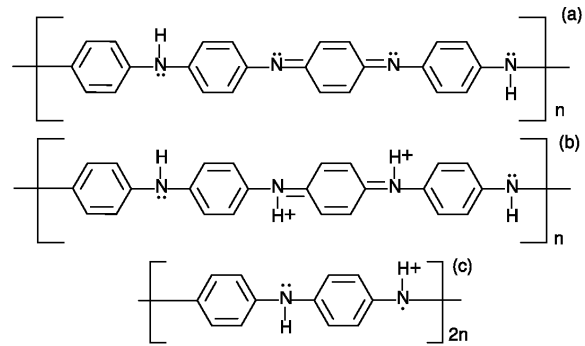}
\end{center}
\caption{Three different forms of PAni: (a) Emeraldine base, and two lattice
models after a doping process, (b) Bipolaron, (c) Polaron }
\label{graph_chem_PANI}
\end{figure}

In spite of its long history, polyaniline (PAni) became a new paradigm for
polymeric conductors as it shows a dramatic increase in conductivity either
by acidic treatment or by electrochemical oxidation. In spite of this fact,
the physical basis of its transport mechanism and of the insulator-metal
transition proved more elusive. Starting from a semiconducting PAni in an
emeraldine base form (Fig. \ref{graph_chem_PANI}-a), protonation leads to an
internal redox reaction that converts it into a metal (emeraldine salt). In
order to account for the highly conducting nature of this doped polymer
there are two well established models that imply two different lattice
arrangements. These are associated to the appearance of two possible charged
defects upon protonation. On one hand, the polaronic lattice (PL), which
describes a lattice of Nitrogen bridged benzene rings that becomes fully
periodic in the case of 100\% of protonation. Even when one of every two
Nitrogens is in the form $N^{+}$ supporting a polaron, the corresponding p$%
_{z}$ electrons form a collective band of Bloch extended states which, being
half-filled, behaves as a metal (Fig. \ref{graph_chem_PANI}-c). On the other
hand, in\ a crystalline bipolaronic lattice (BL) the protonated quinoid
units ($NH^{+}=Q=NH^{+}$) are bridged by three benzene rings. The electron
tunneling between neighbor $NH^{+}$ units, leads to a bonding basic unit
that justifies a bipolaronic description. Natural disorder appears through
the fluctuation of the bridge length (Fig. \ref{graph_chem_PANI}-b). Hence,
while further tunneling between units could be possible, within the standard
wisdom, disorder ensures localized eigenstates that prevent propagation\cite%
{localization}. Galv\~{a}o \textit{et al}. \cite{Galvao-disorder} concluded
that disorder should be an essential ingredient in these systems. They made
molecular orbitals calculations of the electronic structure of PAni chains
which showed that disorder pulls the Fermi energy down through the localized
states of the valence band. Later on, Wu and Phillips \cite{PANI-RDM} agreed
with Galv\~{a}o in the role of the protonation, further showing that induced
disorder can be identified with a Random Dimer Model (RDM) \cite%
{RDM,Libro-Phillips}. By adopting Landauer's view that \textquotedblleft
conductance is transmission\textquotedblright\ \cite{Landauer-Imry-RMP}, the
current \textit{motto} of molecular electronics \cite{MolecularElectronics},
it was proved that the short range order of the RDM produces a set of
delocalized or propagating states \cite{RDM,Schulz,Libro-Phillips}. This
opened the possibility that Fermi energy might lie in a delocalized region.
However, Farchioni \textit{et al}. \cite{farchioni-vignolo-grosso}, by using
an \textit{ab initio} parametrization, made a detailed tight-binding based
study of PAni-HCl comparing the BL and PL models \cite{PANI-Parameters}.
They showed that even when the BL model exhibits extended states, its Fermi
energy is far from the high transmission regions. These ideas seemed to
support the PL as the only PAni emeraldine salt capable of metallic
behavior. Indeed, the observation of Pauli susceptibility on conducting
samples was attributed to extended states in a polaronic lattice \cite%
{Heeger-suscep}. However, first-principle energy stability calculations
point into the opposite direction. A BL is by far the more stable energy
configuration when compared to a PL \cite{Parra-Grosso} or its variants \cite%
{sordo}. A picture that could unify these conclusions is that of segregated
metallic (PL) regions and insulating (BL)\ domains. Transport would be
mediated by hopping between metallic fibers in the polymer backbone \cite%
{islands-model}. However, it is not clear that such structures could give a
lower free energy than a pure BL one. Besides, in this model disordered
islands would constitute the conductance bottle-neck for which a microscopic
description is lacking. Further emphasizing the role of BL, it was recently
suggested that susceptibility experiments could not be used to rule out the
bipolaronic structure from conducting samples. This is because an internal
chemical redox equilibrium between bipolaronic structures and a number of
polaronic defects with Curie susceptibility, should manifest as an overall
susceptibility whose temperature dependence would be indistinguishable from
the Pauli paramagnetism \cite{Petr-equilibrium}. In summary, the early works
associate the conducting state of PAni with periodic order because the
existence of extended Bloch eigenstates is a condition for coherent
propagation.

In order to account for the surprising frequency dependence of the
dielectric constant and of the conductivity observed on conjugated polymers 
\cite{Stafstrom-Epstein-MacDiarmid}, Prigodin and Epstein \cite%
{Prigodin-et-al} suggested a new mechanism of charge transport. They argued
that the metallic state of polymers like PAni is sustained by a granular
picture of transport where metallic islands, separated by amorphous
material, interact through intrachain resonant tunneling events in a quasi
1-D variable range hopping theory. However, after an energy scale analysis,
Martens \textit{et al}. \cite{Martens-et-al, citaGammaTheorico}\textbf{\ }%
arrived to the conclusion that intrachain charge carrier delocalization
should extend over several grains. In consequence, there is some critical
mechanism that governs the formation of truly delocalized states. They
propose a quasi 1-D model of weakly coupled disordered chains with
phase-breaking events that are modeled in the Landauer-B\"{u}ttiker
framework. In this case the 1-D Schr\"{o}dinger wave function picture for a
single chain remains essentially correct with the additions of a finite
lifetime, i.e. decoherence, due to dephasing events.\ Their source can be
multiple, ranging from electron-phonon coupling \cite{Luis-e-ph,Luis-PRB} to
through-space electron-electron interactions between charge fluctuations 
\cite{SSComm-Danieli}, or even a weak interchain coupling \cite%
{Pastawski-UsajQZE}. Increasing the interchain coupling eventually will give
rise to a transition from a quasi 1-D to a fully 3-D behavior as
demonstrated by numerical simulations \cite{Stafstrom, Romer}. However, for
conjugated polymers such as PAni, the interchain charge transfer is weak and
a 1-D model that includes decoherence should be a good approximation. Within
this framework, Martens \textit{et al.} invoke dimensional arguments that
explain the anomalous frequency dependence of the dielectric constant and
conductivity of several polymers. However, their conclusions are based on
estimations of the relevant system quantities.

In this work we attempt to elucidate in detail the nature of conductivity of
polymers by taking the PAni bipolaron lattice structure as the case study.
For that purpose, we use realistic \textit{ab initio} based tight-binding
parameters which can be easily reduced \cite{Levstein-polymer} to the
minimal parametrization of the D'Amato-Pastawski (DP) model \cite%
{Damato-Pastawski}. This provides a simple solution to the otherwise complex
Keldysh formulation of transport \cite{GLBE1,GLBE2}. Indeed, this strategy
was applied before to PAni by Maschke \textit{et al.} and Schreiber \textit{%
et al.} \cite{Maschke1,Maschke2,Schreiber}. However, they focused on already
conducting PAni polaronic chains that are affected by decoherence and/or
interchain coupling. This also sums to the recent efforts in including
decoherent processes in molecular electronics \cite%
{Zimboskaya,Nozaki,Gutierrez-Cuniberti}.
At this point a comment on the meaning of decoherence seems in order since
nowadays this term is mostly used in connection with quantum information
theory, where it denotes the (full or partial) collapse of a pure quantum
state due to usually unspecified interactions with the environment. This
might seem different from the usual language of electronic transport in the
solid state. There, one is used to deal with specific interactions, such as
those with phonons, magnetic impurities or other electrons, which may
involve transitions with given selection rules, e.g. $\left\vert
k\right\rangle \rightarrow \left\vert k+q\right\rangle $. These transition
probabilities are evaluated within a Fermi Golden Rule (FGR). In this
approximation, the coupling with the extra degrees of freedom, the
environment, prevents the interference among the component remaining in $%
\left\vert k\right\rangle $ with that in $\left\vert k+q\right\rangle $.
Thus, with some probability the environment \textquotedblleft
measures\textquotedblright\ an electron in the state $\left\vert
k\right\rangle $ and \textquotedblleft re-injects\textquotedblright\ it
incoherently in $\left\vert k+q\right\rangle $. These processes, which
usually involve some inelasticity, are inherently different from the elastic
scattering with imperfections and impurities that produce the interferences
leading to localization \cite{localization}. Quite often one realizes that,
regardless of the specific selection rule, the relevant role of interactions
is just to provide for decoherence, a mechanism that competes with the
coherent scattering that results in localization. This is precisely the
spirit of the \textquotedblleft local phonon\textquotedblright\ bath or the
fictitious voltage probes that lead to the imaginary site energies
introduced in the DP model \cite{GLBE1,GLBE2}. Thus, the idea is that
decoherence from the system-environment interaction provides a knob that
sweeps transport between a Mott's variable range hopping regime and a
typical 1-D metal \cite{GLBE1,Basko-Aleiner-Altshuler}. Transport in
conducting PAni should occur somewhere at this crossover.

We start in section II by recalling the Landauer formulation for the
conductance and introducing the DP model for decoherent transport. In
Section III we describe the application of this model to the BL structure of
PAni and we show in Section IV the effective transmission results and
associated currents in the non-linear regime. In section V we present the
main conclusion: at room temperature decoherence can account for high
conductance in the strongly disordered bipolaronic lattice. The mechanisms
that can contribute to decoherence are discussed in an Appendix. There, we
asses the possible role of interchain coupling and various forms of
electron-phonon interactions. Moreover, a detailed treatment of these
specific models allows us to enlight their differences and similarities with
the well-known Marcus-Hush theory \cite{Marcus} for vibration-assisted
electron transfer.

\section{Effect of decoherence in the conductance}

The Landauer formulation requires to calculate the transmission of the
system. Assuming that the sample's Hamiltonian is known one must incorporate
explicitly the leads connecting to the electrodes. Besides, to take account
of decoherent interactions with an environment DP included several
phase-breaking fictitious probes \cite{Pastawski-Medina}. Thus, an original
molecular orbital Hamiltonian with $N$ orbitals,%
\begin{equation}
\hat{H}_{S}={\displaystyle\sum\limits_{i=1}^{N}} \left( E_{i}\hat{c}_{i}^{+}%
\hat{c}_{i}+{\displaystyle\sum\limits_{j>i}^{N}} \left[ V_{i,j}\hat{c}%
_{i}^{+}\hat{c}_{j}+V_{j,i}\hat{c}_{j}^{+}\hat{c}_{i}\right] \right) ,
\label{Hamilt-sistema-general}
\end{equation}
becomes an effective Hamiltonian that incorporates the leads and the
interactions with the environment:%
\begin{equation}
\hat{H}_{\mathrm{eff.}}=\left( \hat{H}_{S}-\mathrm{i}\eta~\hat{I}\right) +%
\hat{\Sigma}_{L}+\hat{\Sigma}_{R}+\hat{\Sigma}_{\phi},  \label{Hamiltoniano}
\end{equation}
where $\hat{\Sigma}_{L}=\Sigma_{L}\hat{c}_{1}^{+}\hat{c}_{1}^{{}}$ and $\hat{%
\Sigma}_{R}=\Sigma_{R}\hat{c}_{N}^{+}\hat{c}_{N}$ are, respectively, the
self-energies operators describing the escape to the left and the right
leads obtained through a Dyson equation,%
\begin{align}
\Sigma_{L(R)} & =\frac{V^{2}}{\varepsilon-\left( E_{0}-\mathrm{i}\eta\right)
-\Sigma_{L(R)}}  \label{Self-leads} \\
& =\Delta_{L(R)}(\varepsilon)-\mathrm{i}\Gamma_{L(R)}(\varepsilon),
\end{align}
where $\Gamma_{L(R)}$ results proportional to the escape rate, and hence to
the Fermi velocity, at the $L(R)$ lead. We include the effects of the
incoherent processes in the Hamiltonian, e.g. electron-phonon or through
space electron-electron, simply through an imaginary correction to selected
site eigenenergies 
\begin{equation}
\hat{\Sigma}_{\phi}={\displaystyle\sum_{l}} -\mathrm{i}\Gamma_{\phi}\hat{c}%
_{l}^{+}\hat{c}_{l}.  \label{Sigma-Decoher}
\end{equation}
Hence, $\Gamma_{\phi}$ is an energy uncertainty associated to a decay rate
of the local state at site $l$ described by the FGR. We drop any possible
dependence on $l$ simplifying the description. Since $\hat{I}$ is the
identity operator, $\eta$ can be taken as an infinitesimal imaginary part of
the local energy, $E_{i}\rightarrow E_{i}-\mathrm{i}\eta$, resulting in a
decay to the environment in the same sense as the $\Gamma$'s above.

Given the effective Hamiltonian we have the retarded and advanced Green's
functions in terms of the real energy variable $\varepsilon $:%
\begin{align}
\mathbb{G}^{R}(\varepsilon )& =[\varepsilon \mathbb{I}-\mathbb{H}_{\mathrm{%
eff.}}]^{-1}  \label{Definicion Func Green} \\
& =\mathbb{G}^{A\dagger }(\varepsilon ),
\end{align}%
where $\mathbb{H}_{\mathrm{eff.}}$ may be non-linear in $\varepsilon $ and
non-hermitian for $\eta \neq 0$. The imaginary part of the Green's function
enable to evaluate the local density of states at site $i$ as: 
\begin{equation}
N_{i}(\varepsilon)=-\frac{1}{\pi}\lim\limits_{\eta \rightarrow 0^{+}}\mathbf{%
Im}G_{i,i}^{R}(\varepsilon).  \label{DoS con Func Green}
\end{equation}
In a closed system, this is also obtained diagonalizing the Hamiltonian.

According to the optical theorem, the local density of states, 
\begin{align}
N_{j}(\varepsilon)& =\frac{1}{\pi {\sum_{\beta =L,R,\phi }}\Gamma _{\beta{j}%
}(\varepsilon)}\times  \label{prueba} \\
& \underset{\text{sites}}{{\sum_{i=1}^{N}}}\underset{\text{processes}}{{%
\sum_{\alpha ,\beta ={L,R,\phi }}}}\Gamma _{\beta j}
(\varepsilon)G_{j,i}^{R}(\varepsilon )\Gamma _{\alpha
i}(\varepsilon)G_{i,j}^{A}(\varepsilon ),  \notag
\end{align}
can be cast as a flux between the asymptotic states described by a
generalized Fisher-Lee formula. We adopt the notation $\Gamma _{L}\equiv
\Gamma _{L1}$, $\Gamma _{R}\equiv \Gamma _{RN},~\Gamma _{\phi }\equiv \Gamma
_{\phi i}$, to emphasize that each site can decay through different
processes, e.g. $\alpha ,\beta ~\epsilon \left\{ L,R,\phi \right\} $ are the
possible decay processes taking place at sites~~$i,j~\epsilon \left\{
1,...,N\right\}$. With this notation the generalized transmission
probability results:%
\begin{equation}
T_{\alpha i,\beta j}(\varepsilon )=2\Gamma _{\beta j}(\varepsilon
)G_{j,i}^{R}(\varepsilon )2\Gamma _{\alpha i}(\varepsilon
)G_{i,j}^{A}(\varepsilon ).  \label{Trans de un canal}
\end{equation}%
Within the DP model current conservation is imposed at each site and energy $%
\varepsilon$. This requires an incoherent re-injection of every electron
that has decayed through the $\Gamma _{\phi i}$. This leads to the
evaluation of the kernel $\mathbb{W}^{-1}$, describing incoherent density
propagation, where \cite{Pastawski-Medina}:%
\begin{equation}
W_{i,j}=[(\sum_{k=1}^{N}T_{\phi k,\phi i})\delta _{i,j}+T_{\phi i,\phi
j}(1-\delta _{i,j})].  \label{Matriz de transmitancias}
\end{equation}%
From this we obtain the effective transmission through the sample:%
\begin{equation}
\widetilde{T}_{RL}=T_{RN,L1}+{\displaystyle\sum\limits_{i,j=1}^{N}}%
T_{RN,\phi i}\left[ \mathbb{W}^{-1}\right] _{i,j}T_{\phi j,L1}.
\label{Trans efectiva}
\end{equation}%
The right hand side of Eq. \ref{Trans efectiva} contains two contributions:
the first one represents electrons that propagate coherently through the
sample, the second\ term contains the incoherent contributions due to
electrons that suffer their first collision at site $i$ and their last
collision at site $j$.

Since, $\widetilde{T}(\varepsilon)$ piles up all vertical processes into the
energy $\varepsilon$, one can finally approximate the net current through
the sample by:%
\begin{align}
\mathrm{I} & =\frac{2e}{h}{\displaystyle\int} \widetilde{T}%
_{RL}(\varepsilon)[f_{L}(\varepsilon)-f_{R}(\varepsilon )]\mathrm{d}%
\varepsilon  \label{Corriente} \\
& \simeq\frac{2e^{2}}{h}\widetilde{T}_{RL}(\varepsilon_{F})\mathrm{V}\equiv%
\mathrm{GV,}  \label{corriente-lineal}
\end{align}
where $f_{R(L)}$ is the Fermi distribution function at the right (left) lead
and the factor $2$ accounts for spin degeneracy. Here, we assumed that an
electron with energy $\varepsilon$ coming from the left lead arrives to the
right lead with the same energy, which is only true for very small $%
\Gamma_{\phi}$. The second line is the linear approximation for
infinitesimal voltage $\mathtt{V}$ and temperature, where G accounts for
conductance.

\section{The PAni model}

\begin{figure}[ptb]
\begin{center}
\includegraphics[width=3.0in]{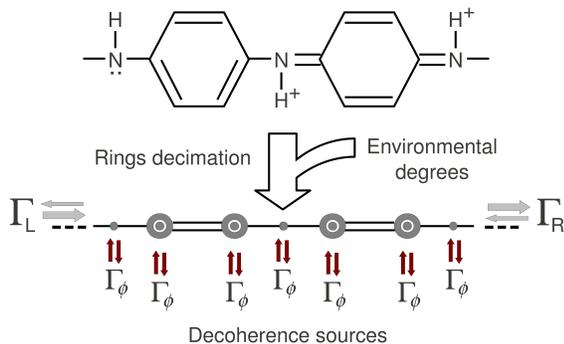}
\end{center}
\caption{Diagrammatic representation of benzenoid rings decimated to obtain
equivalent renormalized units in one dimension. Incoherent channels of
D'Amato-Pastawski model also shown.}
\label{graph_decimacion_anillos}
\end{figure}

We consider a fully protonated BL, which we expect to correspond to the
highly conducting emeraldine salt. By decimation of the benzenoid rings as
it is shown in figure \ref{graph_decimacion_anillos}, we reduce the PAni
emeraldine salt chain to one-dimensional effective system \cite%
{Levstein-polymer}. Each ring is replaced by the proper renormalized sites
at the place of the para-Carbon atoms.

The sample Hamiltonian results:

\begin{equation}
\hat{H}_{S}={\displaystyle\sum\limits_{j=1}^{N}} \left( E_{j}\hat{c}_{j}^{+}%
\hat{c}_{j}+V_{j}\hat{c}_{j}^{+}\hat{c}_{j+1}+V_{j}\hat{c}_{j+1}^{+}\hat{c}%
_{j}\right) .  \label{Hamiltoniano-sistema-explicito}
\end{equation}
When $j=3s+1$ with $s$ positive integer, $E_{j}$ is the Nitrogen p$_{z}$%
-orbital energy and $V_{j}$ is the $\pi$ binding energy (hopping) between
the nitrogen and the para-C p$_{z}$-orbitals. When $j=3s$ and $j=3s-1$ we
have the renormalized parameters for para-C p$_{z}$-orbitals:

\begin{equation}
V_{j}=\dfrac{V_{oo}V_{po}^{2}}{(\varepsilon-E_{o})(\varepsilon-E_{o}-\dfrac{%
V_{oo}^{2}}{\varepsilon-E_{o}})},  \label{Renorm-Hopping}
\end{equation}
and%
\begin{equation}
E_{j}=E_{p}+\dfrac{V_{po}^{2}}{\varepsilon-E_{o}-\dfrac{V_{oo}^{2}}{%
\varepsilon-E_{o}}},  \label{Renorm-site-energy}
\end{equation}
where $E_{o}$ and $E_{p}$ are bare site energies for electrons in the p$_{z}$%
-orbitals of ortho-C and para-C respectively; $V_{oo}$ is the hopping
between ortho-C and $V_{po}$ is the hopping between a para-C and ortho-C. In
this work we use the tight binding parametrization of Vignolo \textit{et al}%
. \cite{PANI-Parameters} for the bipolaron lattice model of base-emeraldine
doped with HCl.

We will consider decoherent sources on effective electron p$_{z}$-orbitals
sites by including a constant imaginary correction to the site energy as in
Eq. \ref{Sigma-Decoher}. This is the most convenient choice for
computational purposes. A first principles calculation of this imaginary
correction is beyond the scope of this work. Its complexity lying on the
multiple effects that must be considered. In the appendix we discuss in
certain detail two mechanisms: interchain tunneling and the effect of
torsional modes on the crucial $\pi $ bonds. However, it is enough to resort
to dimensional arguments based on spontaneous symmetry breaking of the
quantum coherent state \cite{RMP-ContinuousPhaseTransitions}. The reasoning
takes into account that quantum bosonic modes with energies $\leq k_{B}T$
should be occupied by many quanta, indicating that $\Gamma _{\phi }$ should
be of the order of $k_{B}T$ \cite{citaGammaTheorico}. In accordance with
this general framework, Rebentrost \textit{et al}. obtain a comparable
estimation for $\Gamma _{\phi }$ for excitons in photosynthetic complexes
interacting with a phonon bath \cite{Lloyd-et-al}. Indeed, experiments on
DNA strands fits $\Gamma _{\phi }$ of this order using the DP model \cite%
{GammaAjustable}. Thus, we fix an effective $\tilde{\Gamma}_{\phi }$ on each
effective site such that the energy uncertainty per orbital site is $\Gamma
_{\phi }=k_{B}T$. We will see in the next section that small variations of
the precise value of $\Gamma _{\phi }$ have little impact on conductance.

Right and left leads are described by Eq. \ref{Self-leads} choosing $E_{0}=0$
and $V=5e\mathrm{V}$ to observe the appropriate bandwidth of interest $[-10e%
\mathrm{V},10e\mathrm{V}]$. Furthermore, we calculated the Fermi energy by
diagonalizing the exact tight-binding Hamiltonian for different
configurations. For all possible chain arrangements, the Fermi level is
nearly the same, around the average.

\section{Numerical Results}

\begin{figure}[tbp]
\begin{center}
\includegraphics[width=3.0in]{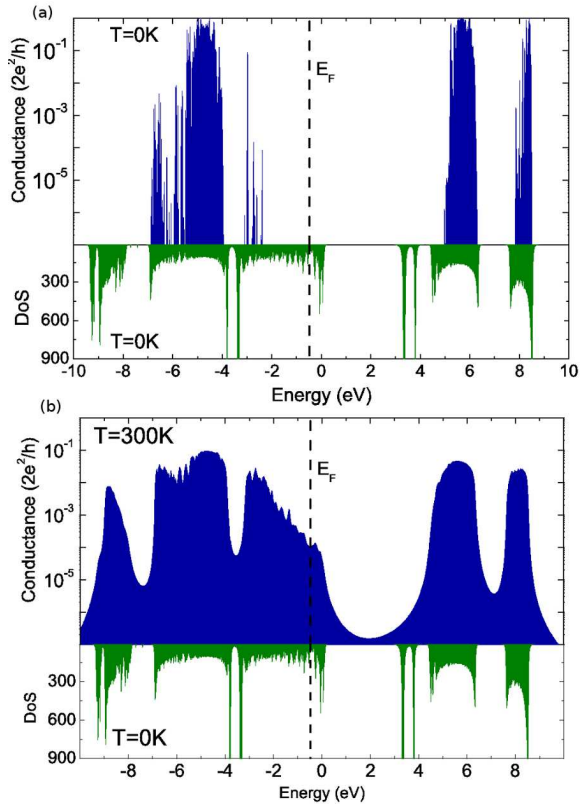}
\end{center}
\caption{Conductance for both: (a) T=0K, and (b) T=300K according to DP
model for a 400 rings long PAni-HCl chain. We also show the T=0K Density of
state for comparison purposes.}
\label{Gr_trans_todas}
\end{figure}

\begin{figure}[tbp]
\begin{center}
\includegraphics[width=3.0in]{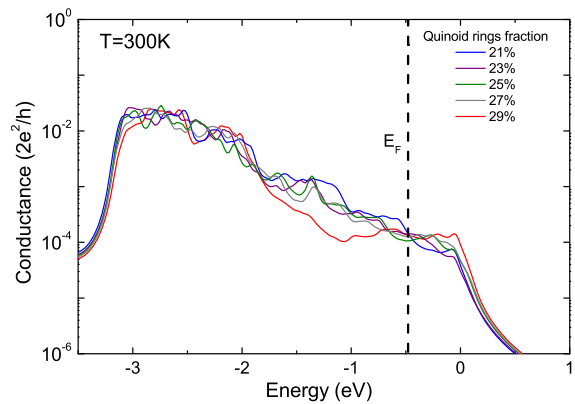}
\end{center}
\caption{Fluctuations in total conductance at T=300K in the main peak around
the Fermi energy with the fraction of quinoid rings along the chain.}
\label{Gr_Quinoid}
\end{figure}

We have performed a detailed analysis of the conductance properties of the
BL model of polyaniline emeraldine salt. Due to the fact that, according to
experimental data, PAni chains seem to have an average length of 400 rings 
\cite{MacDiarmid-Nanillos}, we have taken that number in our numerical
calculations. However, it should be noted that our results do not depend
critically on this parameter. We first calculated the coherent transmission
probability as a function of energy for a set of chain configurations drawn
from the representative ensemble. Results are in full agreement with the
those of Farchioni \textit{et al}. \cite%
{PANI-Parameters,farchioni-vignolo-grosso,Farchioni-renor} and evidence the
mobility edges induced by correlated disorder in this 1-D system \cite{RDM}.
While there is an appreciable density of states at the Fermi energy, it
corresponds to localized states. Indeed, according to Fig. \ref%
{Gr_trans_todas}-a, the Fermi energy is far away from the extended state
region. Fig. \ref{Gr_trans_todas}-b shows the drastic differences in
conductance once that decoherent processes are taken into account.
Conductance at the Fermi energy now becomes appreciable for any
configuration. These results show that metallic transport is possible within
a purely BL model through an environment-assisted transport \cite%
{Lloyd-et-al}. Even within a model of perfectly conducting PL islands
bridged by BL strands, the calculated chains can be taken as representative
of such transport bottleneck. 
One expects that small differences in quinoid ring concentrations would
appear due to natural fluctuations on the\ oxidation degree previous to
doping. In Fig. \ref{Gr_Quinoid} we show the resultant conductance for
various quinoid concentrations and found no significant changes in transport.

\begin{figure}[tbp]
\begin{center}
\includegraphics[width=3.0in]{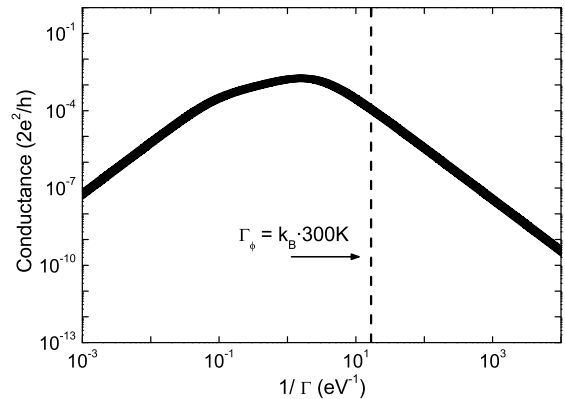}
\end{center}
\caption{Conductance for a 400 rings long PAni-HCl chain as a function of $%
\Gamma_{\protect\phi}$. The value for $\Gamma_{\protect\phi}$ at $T=300K$
also shown in dash line.}
\label{Gr_Conductance_vs_Gamma}
\end{figure}

We also studied the behavior of total conductance as a function of
decoherence rate. In accordance with recent works \cite%
{Lloyd-et-al,Mohseni-Lloyd}, three regimes can be appreciated. Starting from
T=0K, as the temperature rises dephasing events become more successful in
the destruction of localization caused by coherent interference at the Fermi
energy, rising the total conductance of the system. In this regime,
transport rate increases as the energy uncertainty associated with
temperature is increased. However, there is a $\Gamma_{\phi}$ value for
which the conductance is maximal. If the temperature is increased further,
the associated energy uncertainty becomes larger than the terms of the
system Hamiltonian (characteristic hopping and site energies), and the
decoherent process now are able to suppress transport. This is commonly
known in the literature as the quantum Zeno effect. In Fig. \ref%
{Gr_Conductance_vs_Gamma} we show our results for the dependence of total
conductance with $\Gamma_{\phi}$, in which the three regimes above described
are clearly seen. The thermal energy, in this case, lies on an area of great
influence on the total conductance of this system, and therefore decoherent
processes should not be neglected. At room temperatures, the PAni BL is
safely placed in the range of thermally assisted transport, and it is clear
from the figure that small variations in the exact value of $\Gamma_{\phi}$
do not alter the outcome significantly (note the log-log scale).

\begin{figure}[tbp]
\begin{center}
\includegraphics[width=3.0in]{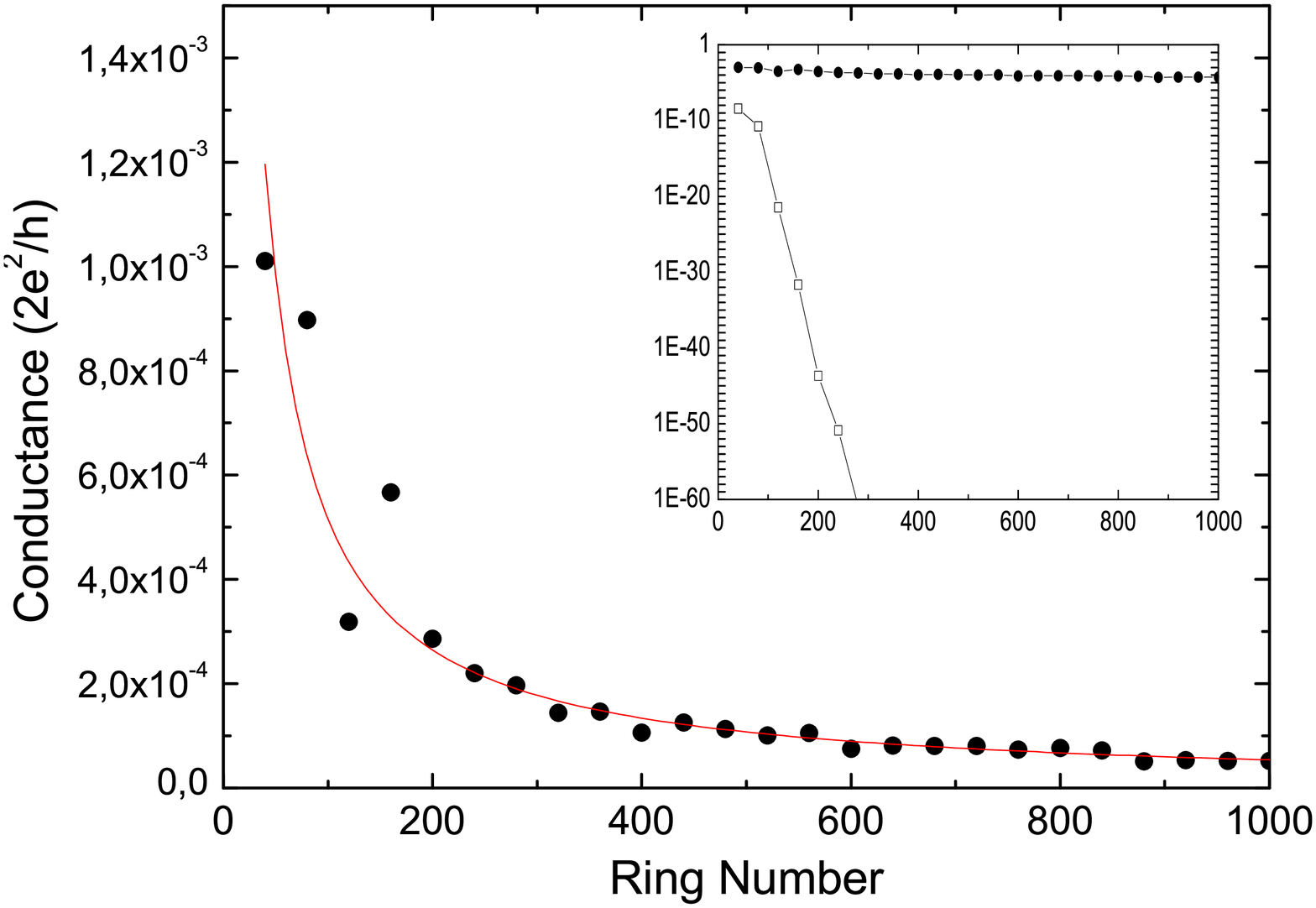}
\end{center}
\caption{Total conductance at T=300K as a function of the chain large. As an
inset, it is shown the scale difference between coherent and total
conductance.}
\label{Gr_Conductance_vs_L}
\end{figure}

We studied the dependence of conductance on the chain length. The results
are shown in Fig. \ref{Gr_Conductance_vs_L}. With the exception of some
fluctuations at short lengths, conductance at T=300K decreases as the
reciprocal of the chain length, as expected for an Ohmic system. The fitting
gives $\mathrm{G}/\left( 2e^{2}/h\right) =1/\left( 20,6N_{R}\right) $, where 
$N_{R}$ is the number of rings of the chain. The log scale in the inset
figure emphasizes the drastic difference between the full conductance and
that restricted to coherent tunneling processes. The coherent conductance
decays exponentially as expected for a one dimensional disordered system 
\cite{exp-decay}. In our case the localization length is small, $\mathrm{G}%
/\left( 2e^{2}/h\right) =22\times10^{3}e^{-0.53N_{R}}$, which implies that
even for very short disordered polymer chains, transport does not take place
unless decoherence processes are involved. Indeed, conductance decays a
factor 1/3 for every two rings. Therefore, in a model of islands as that
mentioned in the introduction, destruction of localization by decoherence
would have a fundamental role. However, our results go further and evidence
that even a fully BL PAni would sustain strong electronic transport.

\begin{figure}[tbp]
\begin{center}
\includegraphics[width=3.0in]{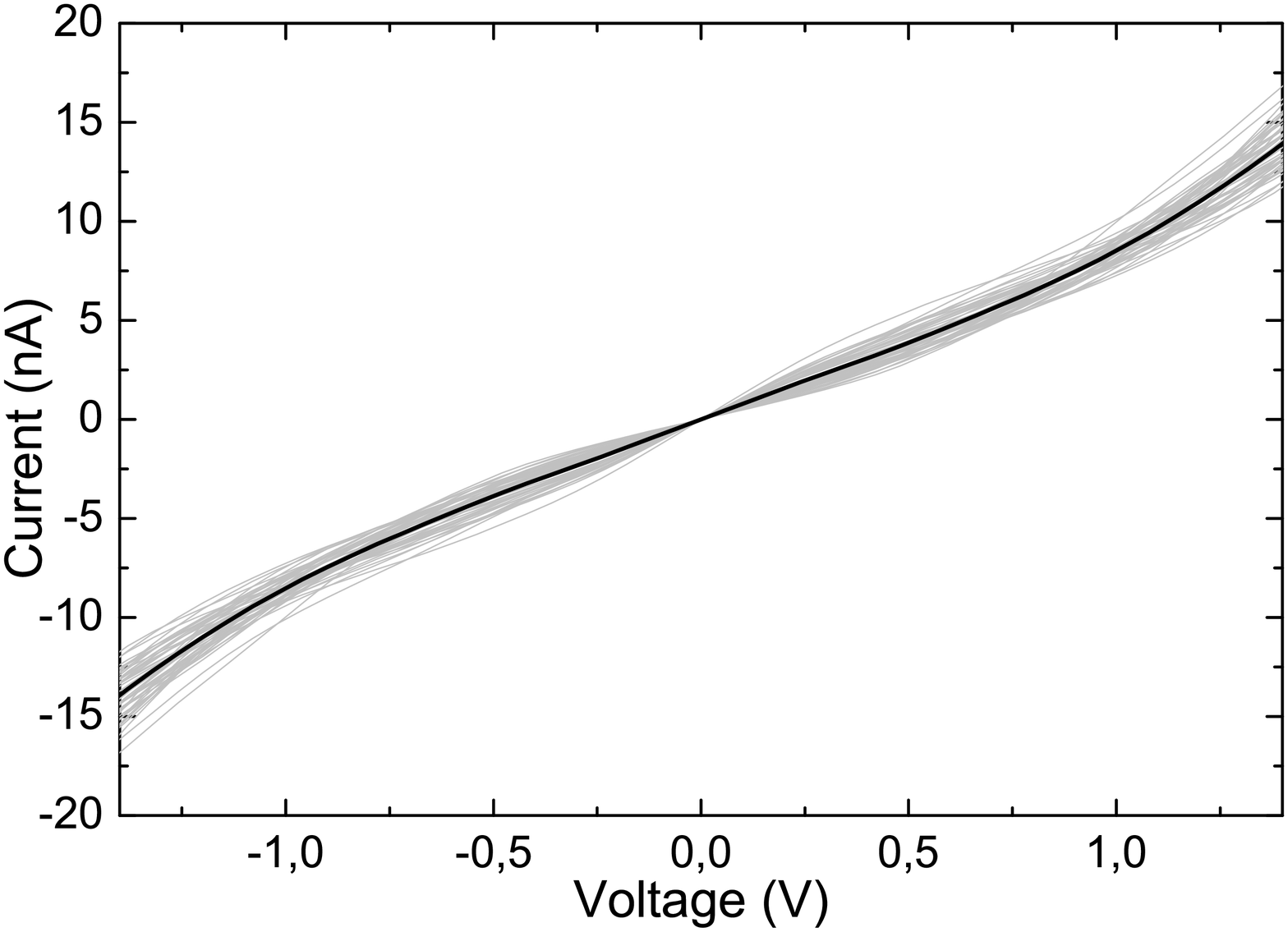}
\end{center}
\caption{Average current (black line) and currents for 50 different
configurations (shadow lines) are shown.}
\label{Gr_corrientesypromedio}
\end{figure}

Fig. \ref{Gr_corrientesypromedio} shows the net current through 50 random
chain configurations of bipolaronic PAni. The current was evaluated by using
Eq. \ref{Corriente} for symmetric voltages at room temperature. As can be
seen, the behavior of different chain configurations is very similar and
close to the linear regime. All chain configurations exhibit an appreciable
conductance with an average value of $7,7$x$10^{-9}\Omega^{-1}$. Taking the
PAni-HCl density as $1,329g/cm^{3}$ \cite{PAni-density} we get a
conductivity of $\sigma=81\Omega^{-1}cm^{-1}$. This result is slightly
higher than the experimental ones, $1\Omega^{-1}cm^{-1}\lesssim\sigma%
\lesssim20\Omega ^{-1}cm^{-1}$ \cite{PAni-conductivity} which is reasonable
because in the calculation of the conductivity we are considering the
conductance of the ideal case of chains directly connected to the leads.

\section{Concluding Remarks}

In this work we have discussed electron conductance in a doped PAni. We show
that the PAni ground state configuration, the BL, has high conductance even
in presence of disorder provided that decoherent processes are included.
This is done without leaving the convenient \textit{a l\`{a}} Landauer
approach by using the generalization introduced by D'Amato and Pastawski 
\cite{Damato-Pastawski} where an effective transmission accounts for
decoherent processes. While our formulation accepts further improvements, it
provides an answer from the robust description of Keldysh formalism within a
minimal parametrization. Roughly speaking, decoherent processes split each
chain into a series of portions whose length is given by the decoherence
length $L_{\phi}$ \cite{Pastawski-Albanesi-Weisz}. These define the
elemental conductivities from which the sample's Ohmic transport builds on.

For many years, it has been assumed that conduction of polyanilines is
inseparably linked to the existence of a polaronic crystalline structure.
However, although our main intention is qualitative, we showed that
decoherent processes are able to give appreciable metallic conduction in the
more entropically favorable bipolaronic lattice. For this system, the
uncertainty of energy associated with thermal processes cannot be neglected
in the study of conductance, since $k_{B}T$\ falls in a region in which the
interplay between incoherent and coherent dynamics results in an increased
efficiency of electron transport. One might then speculate that only when
the thermal energy scale becomes smaller than the Coulomb energy of the
localized states, one would actually start to notice a qualitative
difference with an ideal 1-D metal.

The robustness of the results obtained is evident by noting that they
neither depend on variations in the oxidation degree of PAni prior to the
doping process, nor on the particular arrangement of quinoid rings along the
chain, or on the exact value of the energy uncertainty associated to $\Gamma
_{\phi }$. This justifies the fact that good conducting properties do not
depend much on the purity of the emeraldine base so that small displacements
toward the leucoemeraldine or pernigraniline are acceptable. The evaluation
presented in the appendix show that, even when interchain coupling can
contribute appreciably to conductivity, the coupling between the $p_{z}$
bonds with torsional degrees of freedom is strong enough to provide almost
all the required decoherence. This hypothesis seems consistent with the
experiments that show that adding residues that restrict the torsional
motion would also diminish the conductivity as compared with the unmodified
bipolaronic lattice \cite{polymer-residues}.

We do not attempt to rule out the presence of phase segregation into
metallic polaronic islands and \textquotedblleft
insulating\textquotedblright\ bipolaronic domains. However, these last
strands constitute the bottle-neck where thermal decoherent processes
activate the conductivity. Moreover, our results go further ahead and
evidence that bipolaronic chains can sustain electronic transport by
themselves. In fact, based in our simulations we can estimate bulk
conductivity for these chains and arrive to a remarkably good value as
compared with experimental data.

\section{Acknowledgments}

This work is part of an experimental study of PAni based conductors with the
support of ANPCyT, CONICET, MiNCyT-Cor and SeCyT-UNC for which the
participation of G. A. Monti, P. R. Levstein, R. A. Iglesias and L. J.
Gerbino is acknowledged. Discussions with C. Barbero probed also very useful
for this work. HMP and RABM also acknowledge the hospitality of BIOTECH at
Technical University of Dresden and MPI-PKS where these issues where
discussed with G. Cunniberti, R. Gutierrez and D. Nozaki.

\section{Appendix}

In this section we attempt to elucidate the meaning of $\Gamma _{\phi }$ by
presenting two physically meaningful sources of decoherence which must be
present in our system: interchain coupling and a simple but general model of
electron-phonon interaction. These presentation also allows to compare the
similarities and differences between the decoherence rate and the
electron-transfer rate in the Marcus-Hush model.

\subsection{Interchain hopping}

\begin{figure}[tbp]
\begin{center}
\includegraphics[width=2.9in]{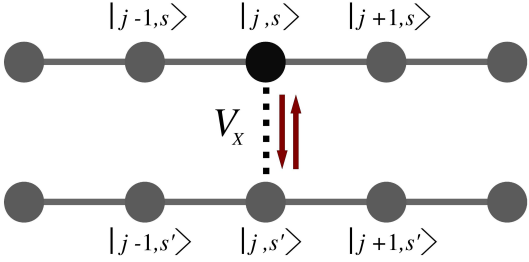}
\end{center}
\caption{Interchain hopping at site $j$. States are written in Dirac
notation including quantum numbers $s$ which label different PAni chains.
This representation illustrates the similarity with Fock-space
representation of the electron-phonon system.}
\label{GR_apen_interchain}
\end{figure}

We start considering the effect of $V_{X}$, an interchain hopping at site $j$%
. Any neighboring chain can act as an \textquotedblleft
environment\textquotedblright\ for an electron at this site. This is because
an electron jumping into a side chain (see Fig. \ref{GR_apen_interchain})
has two options: 1) to escape towards this alternative propagation channel
and never return. This is obviously decoherent as it can not interfere any
longer with the main pathway\cite{Luis-EurophysLett, BoncaPRL}. 2) to return
after having an ergodic walk on the side chain. In this case it is just the
excessive amount of interferences and anti-resonances involved that leads to
a decoherent description \cite{Luis-EurophysLett}. Each node in the plot
corresponds to a multi-chain electronic state. Notice that the interaction
structure looks the same as in the local phonon picture discussed in Ref. 
\cite{Luis-e-ph,Luis-PRB}. Using Eq. \ref{Self-leads}, we have for
self-energy $\Sigma _{j}^{X}$ describing this coupling:

\begin{equation}
\Sigma _{j}^{X}=\frac{|V_{X}|^{2}}{\varepsilon -\left( E_{j}-\mathrm{i}\eta
\right) -\Sigma _{j}}=\left( \frac{V_{X}}{V_{j,j+1}}\right) ^{2}\Sigma _{j},
\label{Self-interchain}
\end{equation}%
where $E_{j}$, $V_{j,j+1}$are site and hopping strength within the chain.
Thus the interchain rate can be expressed as $\Gamma _{\phi }^{X}=\left( 
\frac{V_{X}}{V_{j,j+1}}\right) ^{2}\Gamma _{j}$ where $\Gamma _{j}$ and the
imaginary part of total self energies at site $j$. We may evaluate an
estimate for the typical $\Gamma _{j}$ by disregarding localization and
considering that the side chain is an infinite PAni strand and using the
representative values of $\overline{E}\simeq -0.3eV$ for and site energy and 
$\overline{V}\simeq -3.6eV$ for intrachain $\pi$ bonds. Thus,

\begin{equation}
\Gamma _{j}^{X}=\left( \frac{V_{X}}{\overline{V}}\right) ^{2}\sqrt{\overline{%
V}^{2}-\left( \frac{\epsilon -\overline{E}}{2}\right) ^{2}}.
\end{equation}%
We might wonder which range of values would be required from $V_{X}$ to
yield an energy uncertainty of the order $\Gamma ^{X}\simeq k_{B}T_{R}$
where $k_{B}T_{R}$ stands for room temperature energy. The use of the
discussed values yields $V_{X}<\overline{V}/12$. While we can not ensure
that this is the case, in every site of the PAni chain, uncertainty energy
associated with interchain coupling is not too far below thermal energy and,
therefore, it is it is not negligible. Indeed, there could be sites in which
interchain couplings are stronger, and they would contribute substantially
to the decay of local states.

\subsection{Torsional Phonon coupling}

Certainly, vibrational degrees of motion are natural sources of decoherence.
We analyze them by introducing a simple model for electron-phonon couplings
that enables the evaluation of the corresponding contribution to $%
\Gamma_{\phi}$. From the geometrical inspection of the molecular structure,
it is obvious that torsional strains on benzenoid rings disrupt $\pi$ bonds
between $p_{z}$ orbitals of para-Carbons and Nitrogens. Their overlap
depends on the angle $\theta $ between the orbital axes. As a result, the
corrected hopping energies can be written as $V=V^{0}cos(\theta )\simeq {V}%
^{0}(1-{\theta }^{2}/{2})$. The natural frequency $\omega _{\theta }$ of
this torsional motion determine the vibrational energy of benzenoid rings. A
self consistent description requires that the restoring force $I\omega
_{\theta }^{2}\theta$, written in terms of the moment of inertia $I$ of the
benzenoid ring, should coincide with the net change in the electronic energy
described by the tight-binding model. In this case it yields

\begin{equation*}
V^{0}=I\omega _{\theta }^{2}
\end{equation*}%
leading to $\hbar \omega _{\theta }\simeq 2\times 10^{-2}\mathrm{eV}%
<k_{B}T_{R}$.

In terms of the second quantization operators $\hat{b}=\sqrt{I\omega
_{\theta }/2\hbar }(\theta +\mathrm{i}\dot{\theta}/\omega _{\theta })$ and $%
\hat{b}^{+}=\sqrt{I\omega _{\theta }/2\hbar }(\theta -i\dot{\theta}/\omega
_{\theta \theta })$ we get the perturbation given by the coupling
Hamiltonian:

\begin{eqnarray}
\hat{H}_{el-ph} = & -\frac{1}{4}\hbar \omega _{\theta } \left( \hat{b}^{+}+%
\hat{b}\right) ^{2}\left( \delta _{j^{\prime},j}+\delta
_{j^{\prime},j-1}\right)\times  \notag \\
& {\sum\limits_{j^{\prime}}}\left(\hat{c}_{j^{\prime}}^{+}\hat{c}%
_{j^{\prime}+1}+\hat{c}_{j^{\prime}+1}^{+}\hat{c}_{j^{\prime}}\right)
\end{eqnarray}

\begin{figure}[tbp]
\begin{center}
\includegraphics[width=2.7in]{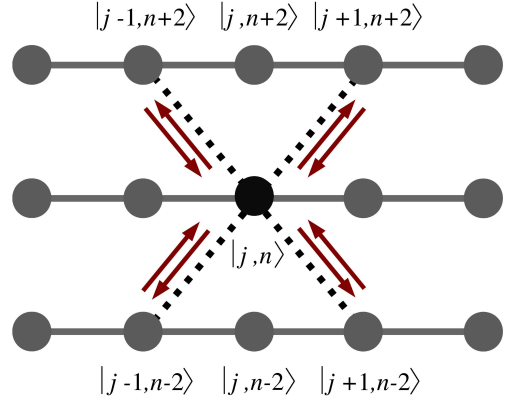}
\end{center}
\caption{Fock-space representation of state $|j,n\rangle $ and its
surroundings. The middle row represents electronic states with $n$ phonons
in the PAni chain. Lower and upper rows represent the same chain but with
different numbers of phonons. Black dotted lines are electron-phonon
couplings.}
\label{GR_apen_fock}
\end{figure}

A Fock-space representation of this interaction Hamiltonian is represented
in Fig. \ref{GR_apen_fock}. Notice the similarities and differences with the
representation of the linear electron-phonon interaction discussed in Ref. 
\cite{Luis-e-ph,Luis-PRB} and the interchain coupling. In the present case,
the effect of the perturbation on the state on a local site $j$ can
evaluated with the FGR:

\begin{eqnarray}
\frac{1}{\tau _{j}(\epsilon )} &=&{\sum\limits_{n}}P(n)\left[ \frac{2\pi }{%
\hbar }{\sum\limits_{j^{\prime },n^{\prime }}}\left\vert \left\langle {j,n}%
\right\vert \hat{H}_{el-ph}\left\vert {j^{\prime },n^{\prime }}\right\rangle
\right\vert ^{2}\right] \times   \notag \\
&&\delta \left[ (\epsilon +n\hbar \omega _{\theta })-(E_{j^{\prime
}}+n^{\prime }\hbar \omega _{\theta })\right] 
\end{eqnarray}%
where $\left\vert {j,n}\right\rangle =\frac{1}{\sqrt{n!}}\left( \hat{b}%
^{+}\right) ^{n}\hat{c}_{j}^{+}$ $\left\vert {\emptyset }\right\rangle $
here $\left\vert {\emptyset }\right\rangle $ is the electron and phonon
vacuum and $n$ label the number of vibrational quantums whose thermal
probability is $P(n)$. In the case of interest, we consider electrons at the
Fermi level, $E_{F}$. Thus, after energy integration and using the thermal
average $\left\langle \left\langle {n}\right\rangle \right\rangle \equiv 
\bar{n}=\sum {P(n)n}$ for the expectation number of $n$, the decay rate
becomes:

\begin{align}
\frac{1}{\tau _{j}}&= \frac{\pi }{16\hbar }(\hbar \omega _{\theta })^{2}\{(%
\bar{n}^{2}+4\bar{n}+2)N(E_{F}-2\hbar \omega _{\theta })  \notag \\
& +2\bar{n}^{2}N(E_{F}+2\hbar \omega _{\theta })+(8\bar{n}^{2}+8\bar{n}%
+1)N(E_{F})\}
\end{align}

We must highlight that the quadratic dependence on displacement in the
electron-phonon interaction is the responsible for the selection rules that
can be appreciated in the last equation. Electrons are allowed to interact
with environment only by absorbing or emitting phonon pairs. This is shown
in Fig. \ref{GR_apen_fock}. However, without much loss of generality that $%
k_{B}T\gg \hbar \omega _{\theta }$, so it is possible to approximate $%
E_{F}\approx E_{F}\pm 2\hbar \omega _{\theta }$ and $\bar{n}\approx \frac{%
k_{B}T}{\hbar \omega _{\theta }}$. As a result,

\begin{equation}
\frac{1}{\tau _{j}}=\frac{\pi }{8\hbar }(\hbar \omega _{\theta })^{2}N(E_{F})
\\
\left[ 12\left( \frac{k_{B}T}{\hbar \omega _{\theta }}\right) ^{2}+12\left( 
\frac{k_{B}T}{\hbar \omega _{\theta }}\right) +3\right]
\end{equation}

The evaluation of the corresponding $\ \Gamma _{\phi }$ becomes trivial in
this high-temperature regime,

\begin{equation}
\Gamma _{\phi }=\frac{\hbar }{2}\frac{1}{\tau _{j}}=\frac{3\pi }{4}%
N(E_{F})(k_{B}T)^{2}.  \label{eq:Gamma_N_kt}
\end{equation}

\begin{figure*}[tbp]
\begin{center}
\includegraphics[width=5.0in]{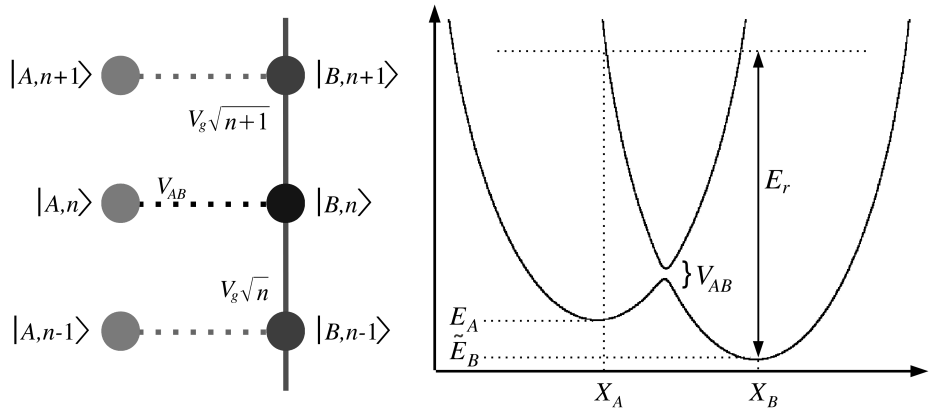}
\end{center}
\caption{Fock-space representation of Eq. \protect\ref{eq:Polaron} and its
corresponding semiclassical picture in terms of vibronic degree of freedom $X
$ as used in the Marcus-Hush model for electron transfer reactions. The
potential surface at the right becomes shifted when represented in the
polaronic basis allowing to define the reorganization energy $E_{r}$. In
this model, the electronic coupling $V_{AB}$ must be small enough to remain
perturbative. Left panel is also used to represent the decoherence by
electron-phonon coupling when $V_{g}\ll V_{AB}$.}
\label{Gr_marcus}
\end{figure*}

Here, it is crucial to notice that for highly localized states the imaginary
self energy results mainly from the decoherent process described above. Thus

\begin{eqnarray}
N(\epsilon ) &\approx &\frac{1}{\pi }\frac{\Gamma _{\phi }}{(\epsilon
-E_{0})^{2}+\Gamma _{\phi }^{2}}  \label{lorenzian} \\
&\approx &1/\pi \Gamma _{\phi }.  \label{lorenzian-height}
\end{eqnarray}

Therefore, from Eq. \ref{eq:Gamma_N_kt}, 
\begin{equation}
\Gamma _{\phi }\sim k_{B}T.  \label{Lorentzian-width}
\end{equation}

Thus, any low frequency modes yielding a quadratic dependence of the
electronic energy on the displacement, which can be more general than
expected, leads to an important consequence: for localized regime, it would
tend to provide an energy uncertainty (decoherence) of the order of the
thermal energy. A similar behavior remains valid if one relaxes the
localization requirement to that of a sharply peaked resonance. A simple
example is a sharp resonance in a one dimensional system \cite{Elena-2009}
in which the local density of states could be written:

\begin{equation}
N(\epsilon )=-\mathrm{Im}\left\{ \frac{1}{\pi }\frac{1}{\epsilon -E_{j}+%
\mathrm{i}\Gamma _{\phi }-\alpha (\Delta -\mathrm{i}\Gamma _{bulk})}\right\},
\end{equation}%
where $\Gamma _{bulk}$ represent the escape to the rest of the tight-binding
chain and $\Delta $ gives the energy shift due the presence of the other
sites. For $\epsilon $ within the band edges, this equation results:

\begin{equation}
N(\epsilon )=\frac{1}{\pi \Gamma }\frac{1+2\alpha \Gamma _{bulk}/\Gamma }{%
\left( \frac{\epsilon -E_{j}-2\alpha \Delta }{\Gamma }\right) ^{2}+\left(
1+2\alpha \Gamma _{bulk}/\Gamma \right) ^{2}},
\end{equation}%
which in the limit of large $\Gamma $ compared with $\alpha \Gamma _{bulk}$,
gives $N(\epsilon )\approx 1/\pi \Gamma $. This limit is achieved at room
temperature whenever $\alpha \ll 1$.

\subsection{Comparison to Marcus-Hush theory}

In the previous section we deemed with the physically relevant situation of
a quadratic interaction with the vibrational coordinate. This is not
conceptually different with the standard linear electron-phonon coupling
used to describe the Franck-Condon effect \cite{Marder} and the
electron-transfer process \cite{Nitzan}. All these physical processes are
contained in a simple Hamiltonian%
\begin{eqnarray}
\hat{H}& = \sum\limits_{j=A,B}{E}_{j}\hat{c}_{j}^{+}\hat{c}_{j}+\hbar \omega
_{0}\left( \hat{b}^{+}\hat{b}+\frac{1}{2}\right) -V_{g}\left( \hat{b}^{+}+%
\hat{b}\right) \hat{c}_{B}^{+}\hat{c}_{B}^{{}}  \notag \\
& +V_{AB}(\hat{c}_{A}^{+}\hat{c}_{B}^{{}}+\hat{c}_{B}^{+}\hat{c}_{A}^{{}}),
\label{eq:Polaron}
\end{eqnarray}%
whose interactions in the Fock space are represented in Fig. \ref{Gr_marcus}%
. The electron transfer problem is best represented resorting to the
polaronic transformation which would diagonalize the Hamiltonian but for the
tunneling described by $V_{AB}$. The essence of an estimation of the
electron transfer rate is a FGR evaluation of the tunneling between the
electronic states A and B in the regime of weak coupling non-adiabatic limit 
$\hbar \omega _{0}$ $\ll $ $k_{B}T,\left\vert V_{AB}\right\vert \ll
\left\vert V_{g}\right\vert $.

\begin{equation}
k_{A\longrightarrow B}=\frac{1}{\tau _{A\longrightarrow B}}=\frac{2\pi }{%
\hbar }\left\vert V_{AB}\right\vert ^{2}\left[ F(\Delta E)\right]
\end{equation}%
where $\Delta E={E}_{A}-\tilde{E}_{B}$, $\tilde{E}_{B}=E_{B}-V_{g}^{2}/\hbar%
\omega_{0}$ and $F(\Delta E)$ is a density of directly connected states
denominated Franck-Condon factor. Thus it satisfies%
\begin{equation}
\int_{-\infty }^{\infty }F(\Delta E)\mathrm{d}\Delta E=1.
\end{equation}
$F(\Delta E)$ is estimated resorting to a thermal average and following the
Marcus original treatment which interprets the transition probability
according to a Landau-Zener formula. Thus%
\begin{equation}
F(\Delta E)=\frac{1}{\sqrt{4\pi E_{r}k_{B}T}}\exp \left[ -\frac{\left(
\Delta E-E_{r}\right) ^{2}}{4E_{r}k_{B}T}\right]
\end{equation}%
where the reorganization energy $E_{r}$ is indicated in the plot.

In contrast to this treatment, in a decoherence problem one focus on
estimating how the electron-phonon interaction degrades the standard
coherent Rabi oscillation\cite{FGR-TLS-horacio}. This describes an electron
jumping forth and back between states B and A and atenuates within a
decoherence time $\tau _{\phi }$. Similarly to Eq. (3.14) of Ref. \cite%
{GLBE2} a FGR evaluation gives:%
\begin{eqnarray}
\frac{1}{\tau _{\phi }} &=&\tfrac{2\pi }{\hbar }\left\vert V_{g}\right\vert
^{2}\left\langle \left\langle (n+1)N(E_{B}+\hbar \omega _{0})+nN(E_{B}-\hbar
\omega _{0})\right\rangle \right\rangle   \notag \\
&\simeq &\frac{4\pi }{\hbar }\left\vert V_{g}\right\vert ^{2}\left[ \frac{%
k_{B}T}{\hbar \omega _{0}}N(E_{B})\right] .
\end{eqnarray}%
where $\left\langle \left\langle {}\right\rangle \right\rangle $ stands for
thermal average. The approximation involves a high temperature limit and
again, the square brackets indicate a density of directly connected states.
As in previous section, the assumption that phonon induced electron energy
uncertainty leads to the self-consistent condition of Eq. \ref%
{lorenzian-height}:%
\begin{equation}
\Gamma _{\phi }=\frac{\hbar }{2\tau _{B}}=\left\vert V_{g}\right\vert \sqrt{%
\frac{2\pi k_{B}T}{\hbar \omega _{0}},}  \label{Decoherence-polaronic}
\end{equation}%
which is valid provided that $\hbar \omega _{0}$ $\ll k_{B}T,\left\vert
V_{g}\right\vert \ll V_{AB}$. Notice that in an ab-intio parametrization of
the tight-binding Hamiltonian, the coupling constant results from evaluating
the dependence of the parameter on the appropriate generalized coordinate,
e.g. $V_{g}=\partial E_{B}/\partial \theta $.

Thus, in both problems, electron transfer in presence of a some
reorganization energy and electron transport with decoherence from a phonon
bath, the Hamiltonian is the same. However, since the calculated observables
are different, the term used as perturbation in the FGR differ. In the first
case the perturbation is the electron jump $V_{AB}$, while in the decoherent
situation the perturbation is the electron-phonon coupling constant $V_{g}$.

\end{document}